\documentclass[english]{elsart}

\usepackage{lineno}

\usepackage[dvips]{graphicx}
\usepackage{epsfig}

\makeatletter
\def\elsartstyle{%
    \def\normalsize{\@setfontsize\normalsize\@xiipt{14.5}}
    \def\small{\@setfontsize\small\@xipt{13.6}}
    \let\footnotesize=\small
    \def\large{\@setfontsize\large\@xivpt{18}}
    \def\Large{\@setfontsize\Large\@xviipt{22}}
    \skip\@mpfootins = 18\p@ \@plus 2\p@
    \normalsize
}
\@ifundefined{square}{}{\let\Box\square}
\makeatother

\begin{document}

%%%%%%%%%
\begin{frontmatter}
\title{The Surface Detector System
\\ of the  Pierre Auger Observatory }
\author[Bariloche]{I. Allekotte},
\author[CBPF]{A. F. Barbosa},
\author[Colorado]{P. Bauleo},
\author[CBPF]{C. Bonifazi},
\author[UTN]{B. Civit},
\author[campinas]{C. O. Escobar},
\author[UTN]{B. Garc\'{\i}a},
\author[bahia1]{G. Guedes},
\author[Bariloche]{M. G\'omez Berisso},
\author[Colorado]{J. L. Harton},
\author[UCLA]{M. Healy},
\author[fermilab]{M. Kaducak},
\author[fermilab]{P. Mantsch},
\author[fermilab]{P. O. Mazur},
\author[fermilab]{C. Newman-Holmes},
\author[bahia2]{I. Pepe},
\author[santiago]{I. Rodriguez-Cabo}
\author[BUAP]{H. Salazar},
\author[Tandar]{N. Smetniansky-De Grande},
\author[Colorado]{D. Warner},
\author[auger]{for the Pierre Auger Collaboration}

\address[Bariloche]{Instituto Balseiro and Centro At\'omico Bariloche (U.N. Cuyo and CNEA, CONICET), 8400
Bariloche, Argentina}
\address[Tandar]{Laboratorio Tandar, Comisi\'{o}n Nacional de Energ\'{i}a At\'{o}mica and CONICET, Av. Gral. Paz 1499 (1650) San Mart\'{i}n, Buenos Aires, Argentina}
\address[UTN]{Universidad Tecnol\'ogica Nacional Regional Mendoza, Mendoza, Argentina}
\address[CBPF]{CBPF, Rua Xavier Sigaud 150, Rio de Janeiro, Brazil}
\address[campinas]{Universidade Estadual de Campinas, Instituto de Fisica, Departamento de Raios Cosmicos, CP 6165, 13084-
971, Campinas SP, Brazil}
\address[bahia1]{Universidade Estadual de Feira de Santana (UEFS),
Av. Universitaria Km 03 da BR 116, Campus Universitario, 44031-460
Feira de Santana BA,  Brazil}
\address[bahia2]{Universidade Federal da Bahia, Campus de Odina, 40210-340 Salvador BA, Brazil}
\address[BUAP]{Benem\'{e}rita Universidad Aut\'{o}noma de Puebla (BUAP), Ap. Postal J-48, 72500 Puebla, Puebla, Mexico}
\address[santiago]{Dpto. F\'isica de Part\'iculas, Universidad de Santiago de
Compostela,  15706 Santiago de Compostela, Spain }
\address[fermilab]{Fermi National Accelerator Lab., Batavia IL, U.S.A.}
\address[UCLA]{University of California, Los Angeles (UCLA), Department of Physics and Astronomy, Los Angeles, CA 90095, U.S.A.}
\address[Colorado]{Colorado State University, Fort Collins, CO 80523, U.S.A.}
\address[auger]{Pierre Auger Collaboration, Av. San Mart\'{\i}n Norte 306, 5613 Malarg\"ue, Mendoza, Argentina}

\ead{ingo@cab.cnea.gov.ar} \ead[url]{www.auger.org.ar;
www.auger.org}

\begin{abstract}
The Pierre Auger Observatory is designed to study cosmic rays with
energies greater than $10^{19}$ eV. Two sites are envisaged for the
observatory, one in each hemisphere, for complete sky coverage. The
southern site of the Auger Observatory, now approaching completion
in Mendoza, Argentina, features an array of 1600 water-Cherenkov
surface detector stations covering 3000 km$^2$, together with 24
fluorescence telescopes to record the air shower cascades produced
by these particles. The two complementary detector techniques
together with the large collecting area form a powerful instrument
for these studies. Although construction is not yet complete, the
Auger Observatory has been taking data stably since January 2004 and
the first physics results are being published. In this paper we
describe the design features and technical characteristics of the
surface detector stations of the Pierre Auger Observatory.
\end{abstract}

\begin{keyword}
Pierre Auger Observatory; high-energy cosmic rays; surface detector
array; water-Cherenkov detectors
\end{keyword}
\end{frontmatter}

%%%%%%%%%

%\linenumbers

\section{Introduction}

Cosmic rays with energies near 10$^{20}$~eV have been a continuing
mystery since Linsley reported the first such event in 1963
\cite{linsley63}. As yet there are no identified sources and no
convincing mechanisms for accelerating particles to these energies.
Interaction with the cosmic microwave background (CMB) constrains
protons of $\sim$10$^{20}$~eV to come from distances not greater
than about 50 Mpc \cite{G,ZK}. Similarly constrained are other
primaries: heavier nuclei lose energy by photo-disintegration and
pair production,  and photons due to pair creation \cite{nagano}.
Furthermore, the flux of cosmic rays at these highest energies is
very low (less than one event per km$^2$  per century per sr), so
that their detailed study requires detectors covering large areas.

The Pierre Auger Observatory was designed for a high statistics,
full sky study of cosmic rays at the highest energies \cite{DR}. It
utilizes an array of surface water-Cherenkov detectors combined with
air fluorescence telescopes, which together provide a powerful
instrument for air shower reconstruction. Energy, direction and
composition measurements are intended to illuminate the mysteries of
the most energetic particles in nature.

On dark moonless nights, air fluorescence telescopes record the
development of what is essentially the electromagnetic shower that
results from the interaction of the primary particle with the upper
atmosphere. The surface array measures the particle densities as the
shower strikes the ground just beyond its maximum development. By
recording the light produced by the developing air shower,
fluorescence telescopes can make a near calorimetric measurement of
the energy. This energy calibration can then be transferred to the
surface array with its nearly 100\% duty factor and large event
gathering power \cite{ICRC-spectrum, ICRC-spectrum07}. Moreover,
independent measurements with the surface array and the fluorescence
detectors alone have limitations that can be overcome by combining
the results of their measurements.

The water-Cherenkov detector was chosen for use in the surface array because of its
robustness and low cost. Furthermore, water-Cherenkov detectors
exhibit a rather uniform exposure up to large zenithal angles and
are sensitive to charged particles as well as to energetic photons
which convert to pairs in the water volume. Their use in surface
arrays was proven to be successful in previous experiments
\cite{haverahpark}.

\begin{figure}
  \begin{center}
    \includegraphics[width=0.9\textwidth]{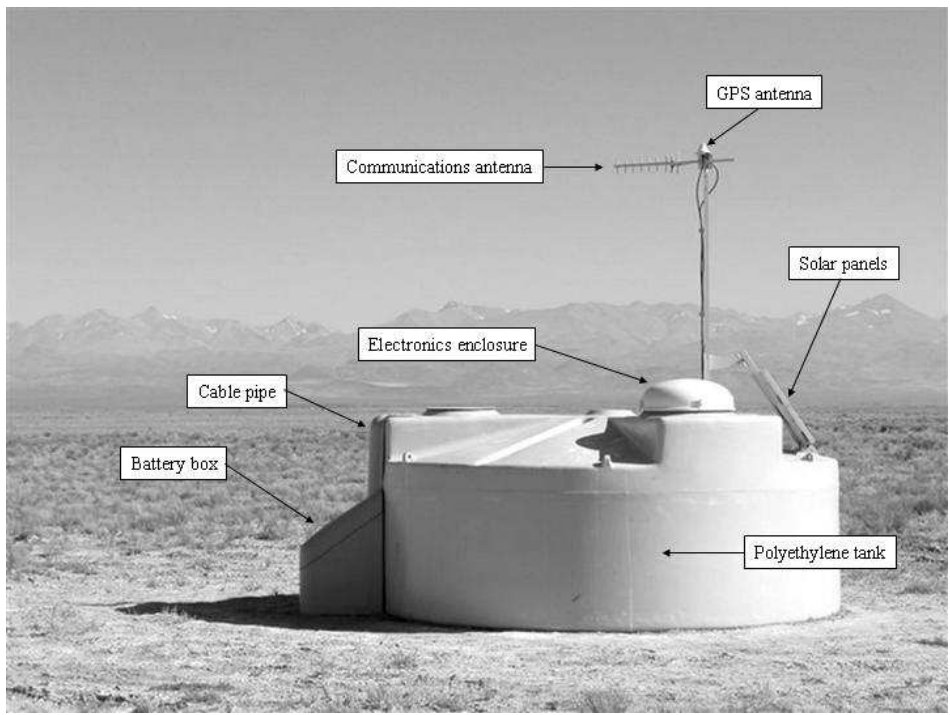}
  \end{center}
  \caption{\label{fig-sd} A schematic view of a surface detector station in the field,
  showing its main components.}
\end{figure}

Each of the 1600 surface detector stations includes a 3.6 m diameter
water tank containing a sealed liner with a reflective inner
surface. The liner contains 12~000~l of pure water. Cherenkov light
produced by the passage of particles through the water is collected
by three nine-inch-diameter photomultiplier tubes (PMTs) that are
symmetrically distributed at a distance of 1.20~m from the center of
the tank and look downwards through windows of clear polyethylene
into the water. The surface detector station is self-contained. A
solar power system provides an average of 10~Watts for the PMTs and
electronics package consisting of a processor, GPS receiver, radio
transceiver and power controller. The components of the surface
detector station are shown in Fig.~\ref{fig-sd}.

In this paper we describe the design features and performance of the
surface detector hardware. This description includes the detector
tanks, liners and accessories and the pure water production, as well
as the most relevant steps for assembly and deployment of the
surface detectors. We conclude with an overview of the technical
performance of the system. The electronics system of the surface
detectors will be described in a companion paper \cite{sde-paper}.

The Southern site of the Auger Observatory, now under construction
in the Province of Mendoza, Argentina, is  over 85\% completed.
Active detectors have been recording events in a stable operation
mode since January 2004 \cite{performance-icrc}.

\section{Design Considerations}

The low event rate of  the highest energy cosmic rays requires an
area large enough to accumulate good statistics in a reasonable
time. By covering an area of 3000~km$^2$ at the Southern Site, the
aperture achieved with the surface array for zenith angles less than
60$^\circ$ will be 7350~km$^2$~sr. By including events with larger
zenith angles, up to 80$^\circ$, the aperture can be increased by
$\sim$30\% \cite{ICRClargezenith}. The detection efficiency at the
trigger level reaches 100\% for energies above 3$\times 10^{18}$~eV
\cite{ICRCacceptance}. This energy is determined from knowledge of
the lateral distribution of showers and the single detector trigger
probability, without recourse to Monte Carlo calculations. The
spacing between the detector stations is the result of a compromise
between cost considerations and the energy threshold (low enough to
ensure a good overlap with existing data.) Other important
considerations are the need for sufficient sampling of the particle
density away from the shower core, and the need for shower front
timing in several locations. A minimum of five stations triggering
at 10$^{19}$~eV allows a maximum spacing of 1500~m on a triangular
grid. At this spacing, approximately 10 stations are triggered by a
nearly vertical shower with an energy of 10$^{20}$~eV. At large
zenith angles the multiplicity of stations triggered increases and
at $\sim$60$^\circ$ it is typically over 20. Differential GPS
systems allow the determination of position and altitude of the
stations with an accuracy of less than 1~m, sufficient for a good
shower reconstruction.

For the installation of this array, the site is required to be flat
for good wireless communications. An altitude between 1000 and 1500
m above sea level is required for optimal development of the shower
in the atmosphere. A large semi-desert area in the west of Argentina
was chosen (35.0$^\circ$ to 35.3$^\circ$ S, 69.0$^\circ$ to
69.4$^\circ$ W) \cite{survey} next to the city of Malarg\" ue . The
chosen site has an average altitude of 1420 m, with detectors
located at altitudes between 1340 m and 1610 m. The site has
suitable infrastructure nearby as well as clear night skies and
minimal light pollution which enables good fluorescence detector
performance.

\subsection{Energy and Angular Resolution and Composition Determination}

The shower energy is obtained by determination of the signal density
at a particular distance (typically 1000~m) from the shower axis.
With the subset of events that the Observatory detects in hybrid
mode (simultaneous measurement with both surface and fluorescence
detectors), the nearly calorimetric energy determination which is
possible with the fluorescence data can be used for an absolute
calibration of the surface detector energy \cite{ICRC-spectrum,
ICRC-spectrum07}.

The signal densities measured with the surface detector array are
affected by fluctuations from different origins: statistical
fluctuations in the measured density, experimental uncertainties on
the shower core position, incidence direction, and large physical
fluctuations in the shower longitudinal development that lead to
shower to shower fluctuations. The sampling fluctuations, which are
dominated by the muon content of the showers, are determined by the
sampling area of the detector. At distances of around 1000~m from
the shower core, the muon flux is of the order of $\sim$ 1~m$^{-2}$
at 10$^{19}$~eV and corresponds to roughly a half of the total
signal, the other half being due to the electromagnetic component of
the shower. Then, with a detector surface of 10~m$^2$ the sampling
error in each detector is below 20\%. For cylindrical detectors,
this corresponds to a diameter of 3.6~m. The statistical uncertainty
(including sampling and reconstruction fluctuations) in the
determination of the signal density at 1000~m from the shower core
is of 10\%~RMS for events with an energy of 5$\times$10$^{18}$~eV
\cite{ICRC-energy, signalaccuracy}.

The direction of the primary is inferred from the relative arrival
times of the shower front at different surface detectors. A weighted
minimization is applied to fit the station triggering times to a
parabolic shower front \cite{tvm}.
A refined determination of the
position of the shower core is obtained by fitting the station
signal densities to the expected lateral distribution. The angular
resolution improves rapidly with energy and zenith angle because of
the greater number of triggered stations. For the surface array
alone, the angular precision is better than 1$^\circ$ for energies
larger than $\sim$ 10$^{19}$~eV \cite{ICRC-Ang, ICRC-Ang07}.

The height of the water-Cherenkov detector is chosen to get a clear
muon signal \cite{ICRC-responsemuons} and optimize the separation of
the muon and electromagnetic contributions to the signal. A vertical
height of 1.2~m of water is sufficient to absorb 85\% of the
incident electromagnetic shower energy at core distances larger than
100~m, and gives a signal proportional to the energy of the
electromagnetic component. Muons passing through the tank generate a
signal proportional to their geometric path length inside the
detector and rather independent of their zenith angle and position.
Each PMT collects in excess of 90 photoelectrons for each vertical
muon \cite{nimcalib}.

\subsection{Physical and Environmental Requirements}

The Observatory will have an operating lifetime of 20 years and must
be designed to survive the expected conditions at the site. The
temperature ranges from -15$^\circ$C to 50$^\circ$ with large
diurnal variation. The outdoor location exposes the detectors to
intense solar ultraviolet radiation and wind of up to
160~km~h$^{-1}$. The detectors must be resistant to floods, rain,
snow, dust, windblown sand and 2.5 cm diameter hail. Material
selection is important because the local soils contain salts which
can be corrosive to some materials. Modest seismic activity should
not damage the detector units. The detector tanks must be robust and
able to support a heavy person on top of the tank as well as to
resist the action of insects, rodents and grazing animals.

The ground on which each detector station is placed must be leveled
to prevent deformation of the tank and the area around the detector
must be cleared of heavy vegetation to avoid damage from bush fires.

\subsection{Design, Development and Production Control}

Each stage in the design, development and production of the surface
detector station was marked by an appropriate technical review.
Subsequent to the preliminary design review, 32 prototype detectors
were built, deployed with standard spacing and operated in a small
Engineering Array \cite{eapaper}. Every design feature of the
detectors, the communications systems and data acquisition was
tested during the two years allocated to the Engineering Array.
Refinements resulting from this period were incorporated into the
baseline design and subjected to a critical design review. A
pre-production run of 100 detectors was then built to qualify the
production process. Production readiness reviews initiated large
scale component production. Assembly and deployment procedures and
associated quality assurance steps were also qualified during the
Engineering Array and pre-production phases.

Individual assembly steps are documented in controlled written
procedures, which are also used for training and guidance of the
staff. A database was developed for the traceability of detector
components and the results of the tests performed on them.

\section{The Tank System}

\subsection{Tanks}

The water-Cherenkov detectors have a cylindrical shape for the water
volume, which is the simplest and least expensive to manufacture.
The top of the tank is rather complex in order to provide rigidity
both for mounting external components such as the solar panels and
for people standing on top of it, and to provide space inside the
tank for the photomultiplier assemblies and cabling. The tanks do
 not exceed 1.6~m in height so that they can be
shipped over the roads within transportation regulations. The beige
tank color is selected to blend in with the natural background of
the site. Although the tank liner and photomultiplier assemblies are
designed for opacity to keep any external light away from the PMTs,
the tank is totally opaque to provide redundancy.

For the manufacture of the surface detector tanks, the technique of
rotational molding (also called ``rotomolding") of high-density
polyethylene was chosen for its low cost, tank uniformity and
because polyethylene meets the requirements of robustness against
the environmental elements.

In the rotomolding process, a predetermined amount of light beige
powdered polyethylene is deposited inside a steel or stainless steel
mold. The inside of the mold has the shape desired for the outside
of the tank. The mold is closed and rotated about two axes
simultaneously inside a 300$^\circ$C oven. The beige powdered
polyethylene melts and forms a coating on the inside surface of
  the mold. Heating and rotation continues until all the powder has been
  deposited on the surface of the mold.
The rotation is briefly stopped and a predetermined amount of black
  powdered polyethylene is put inside the mold, which is immediately re-closed
  and the rotation in the oven continues until all of this powder has been
  deposited on the surface.
Then the mold is removed from the oven and cooled while the rotation continues.
Finally, the mold is opened and the tank removed.

This process, which requires between four and six hours, produces
tanks with a light-beige outer layer of 1/3 of the thickness,
 and an opaque black inner layer
guaranteeing that the tank itself will be opaque. Care in the
manufacturing process results in a nearly uniform wall thickness of
the desired (13 $\pm$ 3)~mm and minimal warping. The nominal weight
of each tank is 530 kg, varying slightly with each manufacturer.
Four companies produced tanks for this project.

Lugs are molded into the tank for lifting it and for supporting the
solar panels. The solar panel bracket lugs are drilled to the
correct diameter after molding and access hatches are cut into the
tank.

The 20-year lifetime of the tanks under outdoor conditions is a
challenging specification. However, discussions with consultants and
experts in the field convinced us that this can be achieved using
high-quality polyethylene resins. To greatly reduce damage due to
ultraviolet exposure, modern polyethylenes contain hindered amine
light stabilizers (HALS). In addition, ultraviolet light is absorbed
by titanium dioxide found in the beige pigment of the outer layer
and by the 1\% carbon black pigment of the inner layer.
 The polyethylene resins used for tank production are prepared in two
stages. The first one is the manufacture of the base resin by
polymerizing selected alkenes with suitable catalysts. This stage of
manufacture also includes the addition of the light stabilizers and
anti-oxidants. The character and quality of the resin are determined
in this stage. The second stage is ``compounding". The polyethylene
resin thus manufactured is melted and the required pigments are
extruded into the resin in such a way that they mix very finely with
the base polyethylene. Other additives, like HALS and antioxidants,
can be mixed in at this stage as well. Then the resin is cooled and
ground into a powder ready for the molding process.

Creep over the 20 years lifetime might also cause the tank to
deform. Creep measurements of samples of our resins and extensive
finite element analysis indicate that creep would not be a problem.
Indeed, no evidence of either creep distortion or ultraviolet
degradation have been observed in any tank, some of which have been
in service for over five years.

\subsection{Hatch Covers and Electronics Enclosure}

\begin{figure}
  \begin{center}
 \includegraphics*[width=0.9\textwidth]{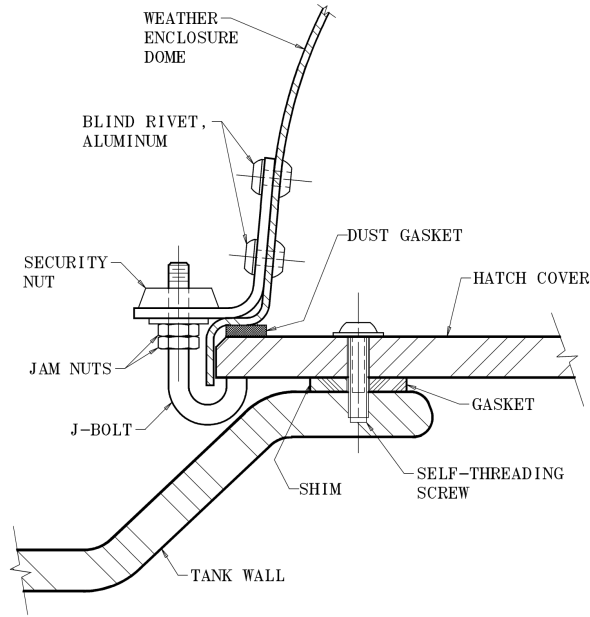}
  \end{center}
  \caption{\label{fig-dome} Details of the hatchcover sealing system and the
  electronics weather enclosure dome attachment}
\end{figure}

As can be seen in Figs.~\ref{fig-sd} and \ref{fig-dome}, the tank
hatches are elevated, to prevent rainwater from accumulating around
the hatchcover and leaking into the tank.

The hatchcover assemblies for the three hatch openings (one large,
560 mm diameter, and two small, 450 mm diameter) consist of
hatchcovers, gaskets, shims between the tank and hatchcover, and
fastening screws. They seal and protect the tank contents, keeping
out light, water and dust. They are easily removable for access to
the tank contents. The large hatchcover is the mounting location for
the electronics and has penetrations for cable feed-throughs.

The hatchcovers are of similar material to the tank itself so that stresses in
the attaching screws are minimized.
Hatchcovers are machined from 12.7~mm high-density, two-layer
polyethylene (HDPE) sheets with beige on the outer
side and opaque black on the inner side. The shape of the hatchcover is
a simple disk with 24 equally spaced holes around the outer edge for
the attaching screws.

The purpose of the shim is to control the spacing between the tank
top and the hatchcover at the location of the gasket, to limit its
amount of compression. The shims (polyethylene) and gaskets (foam
polyurethane) are bonded to the hatchcover using 3M 9472 acrylic
adhesive transfer tape\footnote{3M, St. Paul, Minnesota, U.S.A.,
www.3m.com} which is particularly good at bonding to low surface
energy materials, including polyethylene.

The hatchcovers are attached to the tanks using self threading
screws designed for thermoplastics, identified as Plastite 48-2. The
5.3 mm diameter screws are made from stainless steel and have a
tamper-resistant (pin-in-head) Torx head for increased security.

The detector electronics enclosure is mounted on the large
hatchcover and protected by a weather enclosure, a dome that
provides rain and dust protection and the outer security layer. The
dome can be seen in Fig.~\ref{fig-sd}. The dome itself is spun from
2.3~mm soft aluminum. A foam polyurethane dust gasket is installed
inside the lower lip so that it compresses against the large
hatchcover. The dome is painted  beige to match the color of the
tank. The hold-down system for the weather enclosure consists of a
bracket riveted to the dome, a J-bolt which engages the large
hatchcover, two jam-nuts and one security nut, which can be opened
only with a special tool.

\subsection{Battery Box System}

 Attached to the surface detector station is a rotational molded polyethylene
box containing the batteries
 and charge regulator for the solar power system.
 The battery box, visible in Fig.~\ref{fig-sd}, is placed on the southern side
of the tank, where the tank protects it from direct sunlight to keep
the temperature low and thus increase the lifetime of the batteries.
A polyethylene plate is screwed to the bottom of the box and extends
below the tank to anchor the box, deterring theft or displacement by
large animals. The box has a
 rounded back with radius of curvature equal to the tank radius
  to fit close to the tank. The corners of the
box are rounded to discourage rubbing by cows. The interior of the
box is lined with 50~mm polystyrene foam sheets as thermal
insulation. The top of the box has a slope to deter its use as a
step to get on the tank. The lid is held on with security-head
screws. A protective cover is mounted to the tank to shield the
power system cables that run from the inside of the tank, above the
water level, to the interior of the battery box.

\section{Solar Power System}

\subsection{Solar panels and batteries}

Power for the electronics is provided by a solar photovoltaic
system. The power system provides the required 10~W average power. A
24-V system has been selected for efficient power conversion for the
electronics.

Using the available insolation data for the Auger site, it was found
that a suitable power system can be obtained with two 55~Wp
panels\footnote{Wp is a unit expressed in watts for solar panel
output with a standard solar irradiation applied.}  and two
105~Ampere-hour (Ah), 12 V batteries. Power is expected to be
available over 99\% of the time.
 Even if after long-term operation the capacities of the solar panels and
batteries are degraded to 40~Wp and 80~Ah, respectively, power would
be available 97.8\% of the time.

The batteries\footnote{Model 12MC105, Acumuladores Moura S.A.,
Brazil, www.moura.com.br} selected for the project are a new type of
lead acid battery designed for solar power applications. They have a
selectively permeable membrane and do not require maintenance. Other
lead-acid battery technologies are being considered for replacement
batteries as these wear out.

The charge controller\footnote{Sunsaver SS-10-24V, Morningstar
Corporation, U.S.A., \\
www.morningstarcorp.com} was selected for
robust design and construction, to maximize the lifetime
expectations. An encapsulated, epoxy potted model with robust surge
protection was found in the solar power market. The controller is
pulse width modulated and operates by applying pulses of current of
varying width to the batteries, as their state of charge varies with
battery voltage and temperature. This is considered to be the best
method of charging for maintaining battery efficiency and lifetime.
There have been no observations of electronics interference arising
from the charge controller.

\subsection{Solar Panel Support Brackets and Masts}

The solar panel bracket supports the solar panels and includes the
mast that supports the communications antenna and the GPS antenna.
To optimize light collection in winter time, the solar panels are
installed such that they face North, at an inclination of 55 degrees
with respect to the upward-looking position. The bracket system is
designed to withstand 160 km~h$^{-1}$ winds. The brackets are built
using aluminum 38 mm square tubing with aluminum blind rivets, and
the aluminum alloys used were selected for good corrosion
resistance. The brackets are prepared by cutting, drilling and
riveting most of the assembly in a factory. A few of the rivet holes
are not drilled until the bracket is test-fitted to the tank, so
that the variability in the dimensions from tank to tank is
compensated for. The assembly of the solar panels to the brackets
and of the brackets to the tank is completed before the detectors
are taken out into the field, but the brackets are left in a
collapsed configuration for ease of transportation.  When the
detector is in its final position the panels and mast are raised and
locked in place by a single bolt.

\subsection{Power Cabling}

By mounting the electronics directly on the hatchcover, the length
of cables and the number of connections and feed-throughs are
minimized. The power cables run from the solar panels to the
electronics enclosure and from there through the interior of the
tank to the battery box. They penetrate the large hatchcover and the
tank with light- and water-tight cable feedthroughs. The only cables
exposed to the outside world are the two antenna cables and the
solar power cable coming from the bracket assembly and entering the
electronics enclosure. They are UV protected for outside use. Heavy
gauge wiring was selected for robustness rather than for electrical
resistance considerations.

Sensor cables are installed with the power cables. Voltage of the
individual batteries, their charge and discharge current as well as
the temperatures of the batteries  and the bases of the PMTs are
monitored and registered in 6-minute intervals. The monitoring of
the batteries is also required as the tank power control board is
designed with the capability of setting the local station in
hibernation mode if the voltage drops too low after many days
without sunshine. After a period of prolonged cloudiness, all
stations of the array can be shut down simultaneously rather than
shutting down individual stations, minimizing recovery times and
maximizing data integrity.
 Power system connectors are automotive grade, gold-plated
for long durability in the harsh field conditions.

A grounding rod is driven into the ground at the opening between the
battery box and the tank and connected to the negative terminal to
provide the electronics system grounding.

\section{Liners}

\subsection{Development and Design}

Tank liners are right circular cylinders made of a flexible plastic
material conforming approximately to the inside surface of the
tanks. The liners fulfill three functions: they enclose the water
volume, preventing contamination or the loss of water and providing
a barrier against any light that enters the closed tank; they
diffusively reflect light traversing the water; and  they provide
optical access to the water volume for the PMTs, such that PMTs can
be replaced
 without exposing the water to
the environment.

Three dome windows and five fill ports with screw caps are
hermetically sealed to the liner. The window assemblies allow for
the mounting of the PMTs. The fill ports allow for filling and
venting the tank, as well as providing a window for an LED flasher
used for initial testing.

Although the  tanks provide the primary light barrier for external
light sources, it is necessary that the liners be completely opaque
to act as a secondary protection against small light leaks. Initial
tests were performed to ensure that the laminate and  the seals are
completely opaque against single-photon level light transmission,
i.e., a
 0\% light transmission for light of
wavelengths between 300 and 700~nm, as measured by counting single-photon
detection rates.

Although the mass of water moderates temperature fluctuations, the
temperature range to which the liner is exposed is
from nearly -10$^\circ$C to +50$^\circ$C.
Up to 10~cm of ice could form at the upper surface or sides of the
water volume. The liner is designed to be sufficiently strong and
flexible that it is not damaged by such ice formation. Ice is
prevented from forming near the PMT windows by mounting insulating
rings of polystyrene foam. Strength and flexibility are also
required to withstand the formation of waves up to 15~cm high on the
surface of the water produced by eventual seismic activity. The
Observatory is located in an area rated for moderate seismic
activity and the detector was designed to resist damage from such
activity.

Liner materials require strength, opacity to external light, diffuse
reflectivity of inner surface, sealability, resistance to chemicals
from the environment and to biological activity and minimal
extractables from the material which might contaminate the water
volume enclosed.

\begin{figure}
  \begin{center}
   \includegraphics*[width=0.9\textwidth]{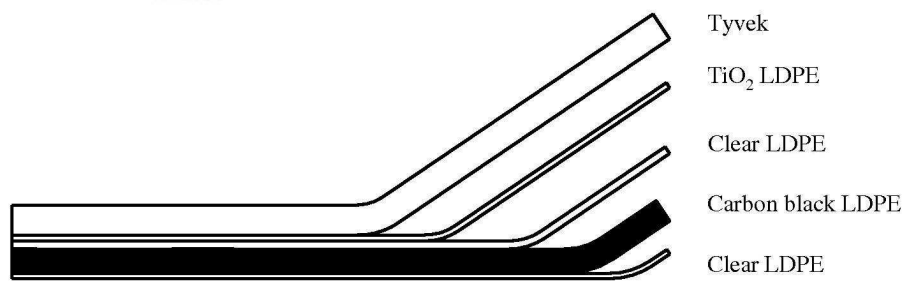}
  \end{center}
  \caption{\label{laminate} A sketch of the laminate,
showing the outer Tyvek\textsuperscript{\textregistered} layer, the
medium TiO$_2$ LDPE layer and the 3-layer with clear LDPE, LDPE with
carbon-black and clear LDPE.}
\end{figure}

The liners are produced from a laminate composed of an opaque
three-layer co-extruded low-density polyethylene (LDPE) film bonded
to a 5.6 mils thick layer of Dupont
Tyvek\textsuperscript{\textregistered} 1025-BL\footnote{E.I. du Pont
de Nemours and Co., Wilmington, Delaware, U.S.A., www.dupont.com} by
a layer of Titanium-dioxide pigmented LDPE of 1.1 mils thickness
(see Fig.~\ref{laminate}). The three-layer co-extruded film consists
of a 4.5 mils thick carbon black loaded LDPE formulated to be opaque
to single photons, sandwiched between layers of clear LDPE to
prevent any carbon black from migrating into the water volume. The
LDPE was metallocene catalized linear low-density polyethylene
(LLDPE) with excellent strength and flexibility.
Tyvek\textsuperscript{\textregistered} 1025-BL was chosen as the
reflective layer due to its strength and excellent diffuse
reflectivity for Cherenkov light in the near ultraviolet
\cite{tyvekrefl,tyvekrefl2}. Tyvek\textsuperscript{\textregistered}
1025-BL is an untreated polyolefin non-woven material, which
minimizes the chemicals available to leach into the water volume. It
is the thinnest of the ``biological grade''
Tyvek\textsuperscript{\textregistered} commercially available, which
simplifies the bonding processes used in manufacturing liners.

Polyolefin film has a strong tendency to pick up electrostatic
charge when unrolled or pulled over a surface and even in a very
clean assembly environment would collect significant dust during the
hours involved in liner assembly. The method for controlling
contamination of the liners centers on minimizing food sources for
microbes by eliminating hair and skin contact with the lamination
and working in a reasonably clean environment. Although the Auger
lamination is not produced in a ``clean room'' environment, the
extruders, lamination machines and slitting machines are all cleaned
prior to production of the Auger lamination, and hair restraints and
gloves are worn during all handling of the film.

\subsection{Assembly and Testing}

Liners are assembled by first manufacturing three separate sections
of laminate and then sealing them together. The separate sections
are the bottom, side strip, and top. The liner top is the most
complex section since it includes the PMT and LED windows and
fill/vent ports. Seals are made by welding the layers together under
pressure with custom designed impulse heat seal machines. The liner
tops were assembled using the same cleanliness procedures as for
laminate manufacture mentioned above. Final liner assemblies were
done in a class 100~000 clean room specially set up for this
project.

All liners were tested for leaks and flaws, and any defects were repaired
before packing and shipping to the site. The same tests were repeated
at the assembly site prior to installation.

\begin{figure}
  \begin{center}
 \includegraphics[width=0.9\textwidth]{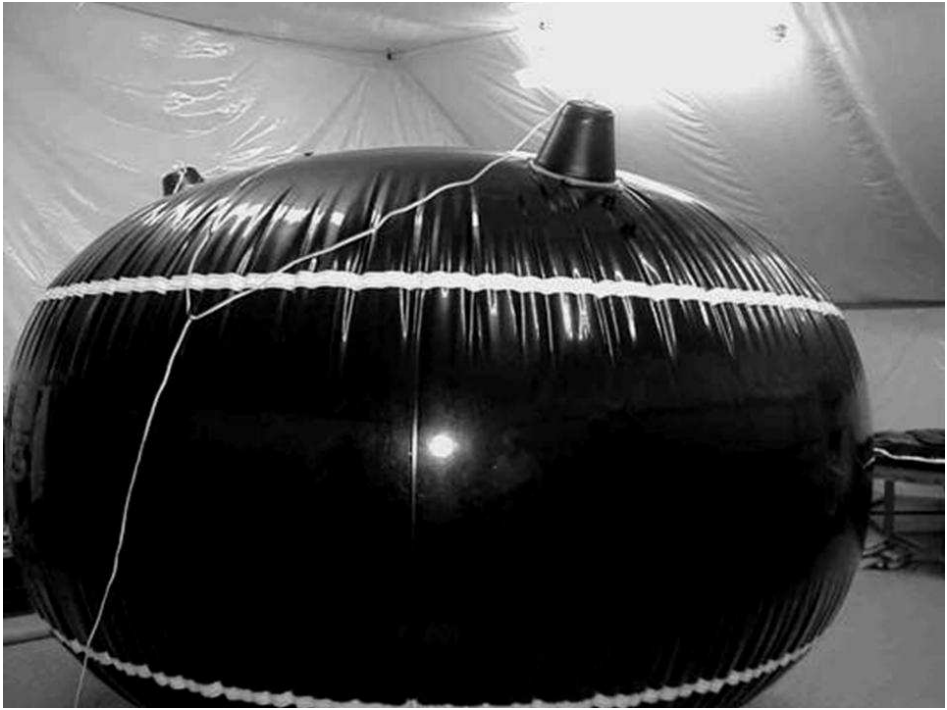}
  \end{center}
  \caption{\label{liner} An inflated liner during testing.}
\end{figure}

For testing,  the liner is inflated to a pressure of 20~millibar
over atmospheric pressure, see Fig.~\ref{liner}. Then all the seals
are tested using a soap bubble solution, looking for visible signs
of bubbling. The liner is then examined in a darkened room with
bright lights covering the window ports such that they only
illuminate the interior of the liner. Any visible light leaking out
from the liner indicates a fault requiring repair. The testing
procedures described above were determined to be sensitive to leaks
smaller than those which  could cause a loss of
 10\% of the detector volume in 20 years.

\subsection{Dome Assembly and PMT Enclosure}

\begin{figure}
  \begin{center}
   \includegraphics[width=0.9\textwidth]{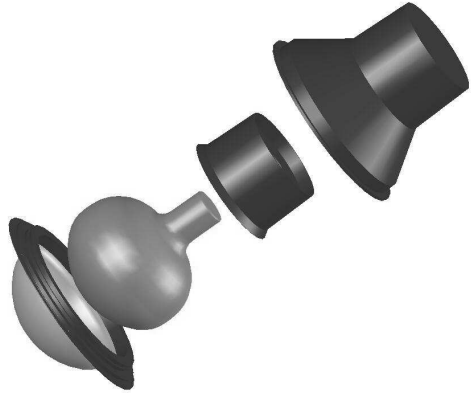}
  \end{center}
  \caption{\label{pmt} Mechanical housing for the PMT
   (top to bottom):  Outer ABS plastic housing;
  insulating plug affixed to the PMT neck that centers the PMT and
  sets the distance of the PMT face to the window;  PMT; flange to
  which the housing is glued with a room-
  temperature RTV; UV-transparent window which is
  fixed to the PMT using a UV-transparent optical coupling.}
\end{figure}

 The PMT enclosures have been designed to
allow the PMT to collect Cherenkov light from the water detector
volume while providing for a cover to shield the PMT from external
light and protecting it from the external environment (see
Fig.~\ref{pmt}).

The foundation of the PMT assembly is an annular, LLDPE flange that
is heat-sealed directly to a hole in the top of the liner using a
custom circular heat impulse welder.  The window through which the
PMT views the water is made of UV-transparent LLDPE.  The windows
are vacuum formed to fit approximately the nominal PMT face. The
window is heat sealed to the flange.  Using heat seals rather than
any adhesive insures that the only material in contact with the
water is polyethylene.  The PMT is protected on the top from light
by an injection-molded ABS plastic cover called the ``fez". For
installation the PMT is indexed with respect to the fez using an
internal polystyrene foam collar that is bonded to the PMT neck. The
variance in the PMT face shape results in a few millimeters
uncertainty on the space between the window and the PMT face, and
that space is filled with 150 ml of the silicone optical compound
GE6136 RTV\footnote{General Electric Company, U.S.A.,
www.gesilicones.com}. Without the optical coupling approximately
half the light from the tank is lost due to total internal
reflection and direct reflection from the interfaces. The fez, with
PMT in place, is sealed to the flange using black GE 123
RTV\footnotemark[6] at the time that the optical seal is made. The
fez has four ports. One port is a light-tight air vent to prevent
pressure buildup due to temperature changes.  The other three ports
are for cable feed-throughs.  These are custom molded two-piece
parts (identical left and right parts) that clip around the cables
and clip to flat annular regions on the fez. There are similar
feed-throughs to pass cables through the bottom of the hatch cover
into the electronics enclosure. Finally there are two annular
polystyrene foam insulation pieces that fit inside the fez to
prevent ice buildup near the PMT. One insulation piece is the same
one that fits on the PMT neck to fix its position with respect to
the fez.  The other fits at about water level and fills the space
between the inside of the flange and the top of the bulb of the PMT
glass. Tests show that ice will form in extreme years on the water
surface, but with the insulation in place it will not stress the
optical coupling or the PMT itself.

\section{Water}

\subsection{Water Quality Specification}

Each surface detector contains 12~000~l of ultra-pure water. The
high water purity is required for two purposes: to achieve the
lowest possible attenuation for UV Cherenkov light, and to guarantee
stability of the water during the 20~years of operation of the
detectors.

For these reasons, the detector water needs to be deionized and
completely free of microorganisms and nutrients. After consulting
experts in pure water production, it was established that the best
achievable water quality requires a water treatment that gives an
output water of resistivity above 15~M$\Omega$-cm.

The production rate of the water plant has to be high enough to
ensure that it can provide water to the detectors at the same rate
as they are deployed. This requirement corresponds to up to
36~000~l/day, which would allow us to fill up to 90~detectors per
month.

\subsection{Water Production}

The pure water required for the surface detectors is produced at a
 plant owned and operated by the Auger Observatory
 and installed at the Central Campus
in Malarg\"ue.

Water is provided both from a local well at 80~m depth and from the
city of Malarg\"ue water network and pumped to a cistern with
60~m$^3$ capacity where it is chlorinated and stored. The water
plant is fed from this cistern. As the quality of the city water is
considerably better than the underground water but more expensive,
the admixture allows an increased production rate and reduced
contamination in the effluents at the lowest cost.

The water purification follows four stages:
\begin{enumerate}
\item  Pre-processing:
\begin{itemize}
\item     Prefiltering, to eliminate particles greater than 40
$\mu$m.
\item Softening with
a resin bed for strong cationic exchange, with regeneration by
sodium chloride, to eliminate Ca$^{++}$, Mg$^{++}$ and Fe ions.
\item    Addition of antiscale solution, to avoid deposit of silicates
  on the reverse osmosis membranes (see below).
\item   Addition of chlorine reducer to eliminate active chlorine.
\item   Microfiltering with two pairs of polypropylene microfilters to
  eliminate particles greater than 5~$\mu$m.
\item   Ultraviolet disinfection with a 254-nm~UV unit (64~W power), to
eliminate microorganisms from the water.
\end{itemize}
\item   Reverse osmosis: A high-pressure centrifugal pump pressurizes
  the water and pumps it to the reverse osmosis unit, consisting of
  two modules in parallel with 4 membranes each and, in series at their output,
 a third module with
4 membranes. Maximum input flow
  is 4500~l/h, with a maximum output of 2800~l/h. The output water
  resistivity is $\sim$ 100~k$\Omega$-cm.
\item Ultraviolet purification: an ultraviolet source of
  185~nm eliminates microbiological residues and removes Total
  Organic Carbon (TOC).
\item  Continuous Electrodeionization (EDI): To achieve the required
  final water quality (resistivity above 15~M$\Omega$-cm), the product
  of the reverse osmosis process is fed to an EDI
  unit\footnote{Model E-Cell MK-1 PHARMA, General Electric Company,
  U.S.A., www.gewater.com}, which consists
  of a set of membranes with cationic and anionic transfer. The
  production capacity of this unit is up to 3400~l/h.
\end{enumerate}

The high-purity output water  is stored in a 50~000 l storage tank.
A recirculation system, which also permits quality improvement
through a mixed resin bed and UV treatment (254~nm, 151~W), can
recirculate up to 5~500 l/h. The pumping system of this
recirculation is also used to fill the transport tank.

The water plant is fully automated. Instruments monitor the working
parameters of the water plant: a chlorine monitor at the entrance of
the reverse osmosis membranes, pressure gauges, flux gauges, flow
meters and resistivity meters. A programmable logical controller
(PLC) records the relevant production parameters.

\subsection{Water Testing and Handling}

The two most relevant parameters that give information about water
quality are its resistivity and biological activity. Resistivity can
be measured continuously at the output of the water plant and in the
storage tank with the instruments incorporated at the water plant.
Resistivity of the water in the transport tank and in the detectors
is determined with hand-held conductivity meters. Although the water
resistivity degrades during transportation, probably due to
absorption of carbon dioxide, tanks in the production phase are
filled with water of 8 to 10~M$\Omega$-cm. During the initial phases
of the project, Engineering Array tanks were filled with water of
less than 1~M$\Omega$-cm quality and they worked as required for
nearly 4 years \cite{eapaper}.

The TOC, i.e., potential nutrients for bacteria, is removed by the
short wavelength UV exposure followed by deionization. The plant
manufacturer specifies that 10 ppb should be achieved. It is not
feasible to measure TOC in the tanks deployed in the field with the
required accuracy, so TOC was only measured a few times at the
output of the water plant, yielding values below 100 ppb.

To determine the biological activity in the water, samples are taken
periodically from the storage tank, the transport tank and the
detectors themselves. These samples are kept in a sterile container
and sent to a biochemists' laboratory to perform the corresponding
analysis and search for aerobe mesophylls, coliforms, faecal
coliforms, coliforms in Koser citrate, yeasts and fungi in Agar
Saboreaud medium. In most of the cases, no biological activity has
been found. Some isolated tanks showed contamination with low
quantities of aerobe mesophylls, identified as being of the genus
``Serratia", which might originate from contamination during
sampling or during sample transportation. In all cases, the
initially detected bacterial contamination was low (below 2000
colony forming bacteria per ml) and these tanks were monitored
periodically and in no case could a large or steady increase in
bacterial activity be detected.

\subsection{Long-Term Stability}

The long-term trends in water quality are tracked using the on-line
tank calibration and monitoring system that is active for every
station and updates every four hours \cite{nimcalib}. In this
application, the time structure of the collected Cherenkov light
produced by through-going muons is recorded to measure the water
purity indirectly \cite{ICRC-water}. Use of the calibration and
monitoring system has the advantage that every station in the array
can be followed and the great quantity of data collected allows some
predictive power even if the measurement lacks the directness of
water sampling.

The charge registered in the fast analog-to-digital converter from
single muons rises rapidly, peaks, and then decreases exponentially.
The decay time depends on the rate of Cherenkov light absorption and
on the reflectivity of the interior of the liner. Measuring the time
constant quantifies the amount of Cherenkov light that is absorbed
in a way that is largely unrelated to the absolute photo-electron
count. An absolute photon count depends on more than just the amount
of absorption in a station. Although it is possible to fit the muon
traces directly and obtain the time constant, this method is
dependent on the precise details of the fitting procedure. For this
reason Q$_{VEM}$ (the total charge deposit by a vertical muon)
divided by I$_{VEM}$ (the signal maximum for a vertical muon) is
used as a substitute for the actual time constant. That ratio can
then be examined as a function of time to search for trends that
have time scales in the range of a few months to several years.

Application of this technique
 allowed us to observe a
decline by 10\% in the Q$_{VEM}$/I$_{VEM}$ ratio in the first
several months after deployment, at which time it reaches an
equilibrium. The origin of this behaviour is still under
investigation. After this the water quality is nearly constant with
a small annual oscillation of less than 1\% in the
Q$_{VEM}$/I$_{VEM}$ ratio, linked to seasonal changes.

\section{Detector Assembly and Deployment}

\subsection{Detector Assembly}

The assembly of the surface detector stations is done in the
Assembly Building located at the Central Campus of the Observatory
in Malarg\"ue. The different components are received and assembled
into a complete detector in this building, which provides workspace
for eight detectors at a time.

When received, tanks are unloaded and inspected. Using a template,
holes are drilled to guide the hatchcover screws and the hatches are
closed to keep water and dirt out of the tank during outdoor
storage. Dimensional measurements, including ultrasonic measurements
of the tank wall thickness, are performed to ensure tank quality.
After cleaning, the tank interior is checked for imperfections that
could damage the liner. Holes for venting, water drain and cable
feed-throughs are drilled into the tank, cables are passed through
the interior of the tank and the liner is inserted and inflated with
filtered air. PMTs are installed and glued to the liner window domes
using optical RTV. Fezzes are mounted over the PMTs to ensure a
light-tight seal. The remaining items are mounted to the tank: the
half-pipe to protect the cables running outside the tank, the solar
panel brackets with solar panels and the electronics enclosure dome.
The liner is kept inflated with air for safe transportation to the
field, and foam pads are inserted between the PMT fezzes and the
hatchcovers to provide cushioning of PMTs during transportation. Six
full time technicians, one foreman and an administrative assistant
(for data entry, inventory and parts receiving and management) can
assemble eight tanks every two days. This includes the assembly of
the solar panel brackets and the preparation of the battery boxes.
PMTs are tested in the Assembly Building after installation, and
serial numbers of the main detector parts are recorded and entered
into the parts management database.

Detector deployment involves survey and site preparation, delivery
of the detector units to their prepared locations, delivery  of
water and installation of the components necessary to complete the
detector. The main challenge for deployment is transportation over
difficult and variable road conditions, particularly with heavy
loads of water. Access to detector locations is affected by seasonal
and daily weather conditions.

\subsection{Site Survey and Preparation}

 Prior to detector deployment,
the ground for each surface detector location is prepared following
these steps:
\begin{itemize}
\item  A contract surveyor
marks the location where each detector is to be deployed with two
stakes oriented north-south at a distance of roughly 10~m from each
other. The surveyor provides the Project with information on the
positioning of both stakes (including altitude) with centimeter
precision, as well as information on ground conditions.
\item   A circular area of 6~m radius is cleared of vegetation. At
locations with pampas grass (``cortadera") or heavy brush, the
circular cleared area is increased to 10 m radius to reduce the
seasonal fire hazard. Local environmental regulations and procedures
are observed.
\item A
central circular area of 2.5~m radius is prepared by clearing it of
stones, roots and other sharp objects and irregularities to avoid
damage to the tank bottom. The ground is leveled to within 3~cm to
avoid overloading the walls of the detector tank and to provide a
uniform water depth and PMT height.
\end{itemize}

The aim was to place all of the detectors on a hexagonal grid of
1500~m spacing. However, for practical reasons, deviations from this
ideal have been inevitable although the median location is within
12~m of the optimum position. Only 4\% of the detector positions are
more than 50~m away from the ideal location, with 0.4\% of the
detectors being displaced more than 100~m. These large displacements
(which have little impact on reconstruction accuracy) were necessary
because of a cultivated area, a riverbed or a swamped and
inaccessible area.

\subsection{Deployment}

 The deployment procedure starts
with loading assembled tanks and transporting them to a staging area
at the site. Tank transport to the staging area is done with flatbed
tractor-trailer trucks carrying four tanks at a time. Staging areas
are selected to be approximately equidistant from the four
deployment locations and in an area where the truck transporting the
tanks can easily maneuver.  An escort vehicle carries other
components for deployment.

 Loading at the Assembly Building is done
with a forklift truck. All tank lifting is performed using the
lifting lugs molded into the top of the tanks along with clevises
and straps. Unloading and further transportation at the site
requires a truck capable of carrying a single tank and equipped with
a hydraulic crane. Such trucks are commonly used for transporting
bricks, drywall, roofing materials and other construction supplies.
While being unloaded and positioned, the tank is oriented such that
the solar panel will face north (5$^\circ$ tolerance) as determined
with a compass. Once the tank is positioned, the battery box is
installed at the south side of the tank and batteries and charge
regulator are installed and connected.

Water is delivered to the detector as discussed in the following
section.  During water filling, the water delivery team
 installs the communications antenna kit and the GPS antenna and mounts
 them to the mast.

Finally, installation of the electronics kit is performed  by a team
of two electronics technicians. The electronics for the detector
station are tested and the detector is commissioned. Contact via a
mobile radio system to a data acquisition technician at the Central
Campus allows the deployment technician to check that the detector
is performing correctly and sending triggers to the central data
acquisition system at the Central Campus before leaving the field.
At this stage the detector is fully integrated into the data taking
system.

\subsection{Water Delivery}

Water is delivered to each detector tank (12~000 l) in one filling,
with a single hose connection. Only a single connection is used in
an effort to prevent contamination by bacteria and/or potential
nutrients.

A water delivery system is composed of two 12~500 l tanks, one
mounted on the back of a truck and one mounted on a trailer. Each
tank has an electrically powered pump,  a gasoline powered
generator, hoses, connections and accessories.
 The trailer is pulled by
the truck on easy access roads and tracks. To access the more
difficult areas, the trailer or the truck are pulled by a large
front-end loader. The front-end loader is also used to even out
irregular roads and to compact the ground in wet areas.

The transport tank system has the following characteristics:
\begin{itemize}
\item  The first transport tank that was acquired for water delivery was made of
fiberglass-reinforced polyester resin with food-grade protective
coatings on the inside.
 The maximum allowable working differential pressure of the tank is 100 cm of
water. For full scale deployment, three additional tanks were
purchased, made of stainless steel (AISI 304 with 2-b sanitary
finishing) as this is more robust to damage in the harsh field
conditions and can be kept clean more easily. Two of these tanks
were mounted on trucks and two on trailers.
\item   A
0.2 $\mu$m bacteriological filter is connected to the air inlet of
the tank to filter the air that is sucked into the transport tank as
the water is pumped out. A valve is installed below the filter, to
ensure that the water cannot splash the filter during truck
movements because the filter has a very high pressure drop when wet.
The valve is opened when the tank is being emptied to allow inflow
of air.
\item   Each tank has a manway to allow access for cleaning. A
pressure relief valve is installed at the manway to avoid damage to the tank by
overpressure during filling.
\item   There is a transparent plastic window on
the tank for direct visual inspection.
\item   A 50 mm hose and associated valves
are installed to transfer the water from the transport tank to the
detector. The end of the hose is connected to a bayonet that has a
valve to regulate water flow and a freely rotating cap that can be
screwed to the liner opening. During transportation, the bayonet is
protected with a stainless steel scabbard which can be screwed to
the bayonet with an airtight seal.
\item   The electrically driven stainless
steel centrifugal pump installed to transfer the water has a
capacity of 120-300 l~min$^{-1}$.
\item   All accessories in contact with water are stainless
steel with a sanitary finish to prevent corrosion and formation of bacterial
colonies.
\end{itemize}

The recirculation system of the water plant is used to fill the
transport tank. This allows a flow of 12~500 l in 50 minutes. Before
filling the tank the water conductivity is tested with a hand-held
conductivity meter. To fill the detectors, hatch covers are removed
after cleaning the tank surface, one liner cap is opened and the
bayonet, after being rinsed thoroughly, is introduced into the liner
and screwed to the liner opening, and the pump is turned on. A
second liner port is opened for air release. The filling of the
detector takes approximately 45 minutes. The height of the water
column is determined by measuring the height of the tank and
subtracting the height of the water level, measured from the top of
the tank. This gives a precision of 1-2 cm. The level is measured at
different hatch openings to avoid systematic errors due to any
possible tank tilt.  After filling, any remaining air is pumped out
of the liners with a vacuum cleaner. Once deployed, water level
measurements can be obtained from the slope of the charge histogram
from single muon tracks \cite{ICRC-water, ICRC-waterlevel}.

After pauses in water deployment of five to six days, the transport
tanks and all accessories are cleaned and disinfected, and filters
are checked and replaced as required. Cleaning is done with
detergent, bleach and a very thorough rinse.

\section{Maintenance and Operation}

%\subsection{Performance, Maintenance and Operation}

As of September 2007, more than 1400 surface detector stations are
operational. Typically more than 98.5\% of the stations are
operational at any time. The technical staff distributes its time
between deployment of new detectors and maintenance and repair of
down stations.

Only seven liners were observed to leak shortly after installation.
In these cases, which constitute the worst failure mode, the tank is
emptied and brought back to the Assembly Building for replacement of
the interior components.

There have been very few instances of human interference with the
surface detectors. During 5 years of operation, only 12 solar panels
have been damaged and two have been removed (both from locations
along side a paved road).

Solar power system parameters are recorded and analyzed using the
central data acquisition system. Failures are treated on an
individual basis. Monitoring software for the solar power system has
been developed to make this monitoring routine for operating
personnel and scientists either on shift at the site or elsewhere by
internet access. The lifetime of batteries is estimated to be four
years. The batteries will be monitored along with the rest of the
solar power system. The condition of the batteries can be determined
from the data (voltages, currents, and temperatures) that are being
monitored and the weak batteries can therefore be identified weeks
or even months before complete failure occurs. Batteries can then be
scheduled for replacement by the routine maintenance process.

\section{Conclusions}

In conclusion, with over 1400 commissioned detectors in the field,
some of which have already been operational for over five years,
much insight on their performance has been gained. All components of
the above-described detector hardware have fully met our design
expectations. The design has proven sufficiently robust to withstand
the adverse field conditions and failure rates are less than
expected. Data taking is ongoing and the first scientific results
have already been published. The physics performance has met or
exceeded all of our requirements
\cite{performance-icrc,ICRC-energy,ICRC-Ang,ICRC-Ang07}.

\section*{Acknowledgements}

The successful installation and commissioning of the Auger Surface
Array would not have been possible without the strong commitment and
effort from the technical and administrative staff in Malarg\"ue.

We are very grateful to the following agencies and organizations for
financial support: Gobierno de la Provincia de Mendoza, Comisi\'on
Nacional de Energ\'\i a At\'omica, Municipalidad de Malarg\"ue,
Fundaci\'on Antorchas, Argentina; the Australian Research Council;
Funda\c{c}\~{a}o de Amparo a Pesquisa do Estado de S\~{a}o Paulo,
Conselho Nacional de Desenvolvimento Cient\'ifico e Tecnol\'ogico,
Funda\c{c}\~{a}o de Amparo a Pesquisa do Estado de Rio de Janeiro
and Financiadora de Estudos e Projetos do Ministerio da Ciencia e
Tecnologia, Brasil; Ministry of Education, Youth and Sports of the
Czech Republic; Centre National de la Recherche Scientifique,
Conseil R\'{e}gional Ile-de-France, D\'{e}par\-te\-ment Physique
Nucl\'{e}aire et Corpusculaire (PNC-IN2P3/CNRS), Departement
Sciences de l'Univers (SDU-INSU/CNRS), France; Bundesministerium
f\"ur Bildung und Forschung, Deutsche Forschungs\-gemein\-schaft,
Helm\-holtz-Ge\-mein\-schaft Deut\-scher Forschungszentren,
Fi\-nanz\-mi\-nis\-te\-rium Baden-W\"urttemberg, Ministerium für
Wissenschaft und Forschung Nordrhein Westfalen, Ministerium f\"ur
Wissenschaft, Forschung und Kunst Baden-W\"urt\-temberg, Germany;
Istituto Nazionale di Fisica Nucleare and Ministero dell'Istruzione,
dell'Universit\'a e della Ricerca, Italy; Consejo Nacional de
Ciencia y Tecnolog\'{i}a Mexico; Ministerie van Onderwijs, Cultuur
en Wetenschap, Nederlandse Organisatie voor Wetenschappelijk
Onderzoek, Stichting voor Fundamenteel Onderzoek der Materie,
Netherlands; Ministry of Science and Higher Education, Poland;
Funda\c{c}\~{a}o para a Ci\^{e}ncia e a Tecnologia, Portugal;
Slovenian Ministry for Higher Education, Science, and Technology and
Slovenian Research Agency; Comunidad de Madrid, Consejer\'ia de
Educaci\'on de la Comunidad de Castilla La Mancha, FEDER funds,
Ministerio de Educaci\'on y Ciencia, Xunta de Galicia,
 Spain;
Science and Technology Facilities Council (formerly PPARC), UK; the
US Department of Energy, the US National Science Foundation, The
Grainger Foundation, U.S.A.;  UNESCO and the ALFA-EC in the
framework of the HELEN Project.

\end{document}